# The Future will be Different than Today: Model Evaluation Considerations when Developing Translational Clinical Biomarker

Yichen Lu*
Jane Fridlyand
Tiffany Tang*
Ting Qi
yl002013@uw.edu
fridlyand.jane@gene.com
tiffany.tang@berkeley.edu
qi.ting@gene.com
Genentech Inc.
South San Francisco, CA, USA

Noah Simon
Department of Biostatistics,
University of Washington
Seattle, WA, USA
nrsimon@uw.edu

Ning Leng
Genentech Inc.
South San Francisco, CA, USA
leng.ning@gene.com

**Abstract**

Finding translational biomarkers stands center stage of the future of personalized medicine in healthcare. We observed notable challenges in identifying robust biomarkers as some with great performance in one scenario often fail to perform well in new trials (e.g. different population, indications). With rapid development in the clinical trial world (e.g. assay, disease definition), new trials very likely differ from legacy ones in many perspectives and in development of biomarkers this heterogeneity should be considered. In response, we recommend considering building in the heterogeneity when evaluating biomarkers. In this paper, we present one evaluation strategy by using leave-one-study-out (LOSO) in place of conventional cross-validation (cv) methods to account for the potential heterogeneity across trials used for building and testing the biomarkers. To demonstrate the performance of K-fold vs LOSO cv in estimating the effect size of biomarkers, we leveraged data from clinical trials and simulation studies. In our assessment, LOSO cv provided a more objective estimate of the future performance. This conclusion remained true across different evaluation metrics and different statistical methods.

## 1 Introduction

Personalized healthcare has become one of the fastest growing research areas in medical research. Ideally, patients could expect to have tailored treatment plans based on their individual profile. The key to enabling such customization lies in identifying diagnostic biomarker signatures that can determine if a patient has a specific condition to benefit from a certain type of treatment. While we have seen a few rising biomarkers on the horizon, many of them are far from being perfect with reliable performance in different populations and being generalizable across indications. [1][2].

The path to finding a good and robust biomarker can be tricky. Often we risk having biomarkers that perform well on legacy trials but cannot be transferable to heterogeneous future trials due to our model overfitting to our training "context". This context, which includes characteristics of the patients, the disease, and even how the biomarker is measured, will often be different between legacy trials and future trials. If we want to understand how well the model will transfer, it is of interest to use cross-validation (cv) methods that directly train and test in different contexts.

For conventional cv methods such as K-fold, the sample splitting mechanism indicates the assumption behind K-fold cv: the legacy and future trials will resemble each other. Nevertheless, heterogeneity across trials can violate this assumption, leaving us an overly optimistic estimate of biomarker's effect size at hand. Within-study cv, meta-analysis based on within-study cv, and cv on pooled data after batch effect correction may also have the same problem, as the random split step implies that legacy and future trials come from the same probability generating distribution. The extent of overestimation worsens as the future trial differs more from the legacy trials used for training.

To correct the over-optimism in our estimation, we recommend incorporating "fit-for-purpose" cv, in which the training and validation folds are set up to mimic the difference between the learning data set and the future data set. "Fit-for-purpose" cv is not a brand new concept: we already saw a few previous cases where it was used to validate biomarkers and assess the performance of a model[3][4]. We mean to emphasize the importance of accounting for heterogeneity while training a model and correcting over-optimism when

---

we estimate its future performance, and "fit-for-purpose" cv plays a key role in finding such generalizable signatures.

For this paper, the objective is to show that the leave-one-study-out (LOSO) cv, one example of the "fit-for-purpose" cv, can more objectively estimate test error of a biomarker on unseen trials compared to commonly used K-fold cv. We start with a case study example of developing a gene-expression based biomarker to identify patients with higher probability of response to the anti-PD-L1 treatment atezolizumab. We introduce evaluation metrics for the models and compare the performance of the K-fold and LOSO cvs. The next section describes a simulation study. It examines the performance of the two cv methods on simulated trials with shifting structures, and expands the type of performance metrics used for comparing models. The last part provides summarization of the findings and discussions.

## 2 Case study

### 2.0.1 Background.

The case study illustrates a clinical trial scenario for learning biomarker signatures where multiple candidate genes were evaluated using legacy trials. The goal is to identify a biomarker (a single feature or a combination of features) that can classify patients receiving atezolizumab into two groups with higher or lower likelihood to respond to the treatment for a future trial.

Patient-level data from four legacy trials (POPLAR, OAK, IMpower150, IMpower131) were extracted. All trials include patients with non-small cell lung cancer, but the population differs in disease stage and the inclusion/exclusion criteria [5][6][7][8]. This case study emulates the decision making process with objective response rate (ORR) before the readout of IMpower131 by incorporating insights generated from the three legacy trials. For ORR, responder is defined as patients with their best confirmed overall response by investigator of Complete Response (CR) or Partial Response (PR), and non-responder otherwise.

All four trials compare at least one atezolizumab-alone or atezolizumab combination intervention to other control arms. In this case study, we mimic a single arm trial by only engaging the treatment (atezolizumab) arm to build and evaluate biomarkers in grouping patients into groups with better or worse clinical outcome. This approach is sometimes used for (1) prognostic biomarker identification when the biomarker is known to be prognostic in the standard of care, (2) predictive biomarker when no standard of care exists, or no/very few patients respond to the standard of care. In the latter case, the response rate difference between two biomarker high and low groups in the treatment arm can be directly translated to treatment effect differences.

Following the same approach, we treat each atezolizumab arm as a single trial and put together four single-arm cohorts from three legacy studies, one each from POPLAR (n=144) and OAK (n=613), atezolizumab + carboplatin + paclitaxel cohort (n=402) and atezolizumab + carboplatin + paclitaxel + bevacizumab cohort (n=400) from IMpower150, and one single-arm cohort (atezolizumab + carboplatin + nab-paclitaxel from IMpower131) as the "future" trial (n=325). Such simplification is still illustrative of the process of building and validating biomarkers.

RNASeq gene expression at baseline and demographic information were collected among all cohorts. The final dataset included five basic baseline characteristics including age, sex, race, ethnicity, and country.

### 2.0.2 Method.

To find biomarker signals, we leveraged the four atezolizumab cohorts from POPLAR, OAK, IMpower150 to train and evaluate models with either of the cv methods: K-fold or LOSO. For K-fold cv, patients were randomly partitioned into four equally sized folds so each fold included patients from different cohorts. For LOSO cv, each cohort acted as a training fold and each fold only included patients from one single-arm atezolizumab group. During the cv process, a statistical model was built on three folds (training set), predicted patients' response on the other held-out fold (validation set), and calculated the corresponding performance. We repeated these steps four times so each cohort was held out once as the validation set and the average of the three measurements was the estimate of the true performance of the biomarker model on the "future" trial.

While the outcome of the "future" trial was not available in the decision making stage, the case study had the "future" trial (IMpower131) unblinded in the end so we were able to calculate the true performance of the biomarker using patients' clinical outcomes and evaluate whether K-fold and LOSO cv give an objective prediction.

To show that the overfitting phenomenon is agnostic to model type, we deployed three statistical models: Lasso regression, random forest and gradient boosting machine. These biomarker-based prediction models assigned a predicted probability of response-to-treatment for each patient, and we evaluate the biomarker performance based on how well the model split the population into biomarker positive and negative groups. We used three metrics: AUC, calibrated difference in ORR (ΔORR), and calibrated classification accuracy[9][10]. AUC examined model performance by looking at all possible classification thresholds that divide patients into biomarker positive (predicted probability >= threshold, response to atezolizumab) and negative (predicted probability < threshold, no response) groups. Calibrated ΔORR and classification accuracy split patients into biomarker positive and negative cohorts based on a specific threshold that is "calibrated" through dichotomizing the predictions using the future study. Since we aimed at a prevalence rate of 50%, we used the median predicted probability from the "future" trial to categorize patients.

The Future will be Different than Today: Model Evaluation Considerations when Developing Translational Clinical Biomarker

### 2.0.3 Results.
Figure 1 shows the estimated AUC and ΔORR between biomarker high and low groups, using K-fold or LOSO cv schemes. The K-fold cv overestimates the true AUC and ΔORR, and the magnitude of over-optimism is not neglectable. For instance, with the Lasso regression model, K-fold cv and LOSO estimated the mean ΔORR to be 0.34 and 0.18, respectively, while the true ΔORR was 0.13. In contrast, LOSO cv gives a closer estimate, confirming that it reduces more over-optimism over standard K-fold cv, regardless of the predictive method we chose.

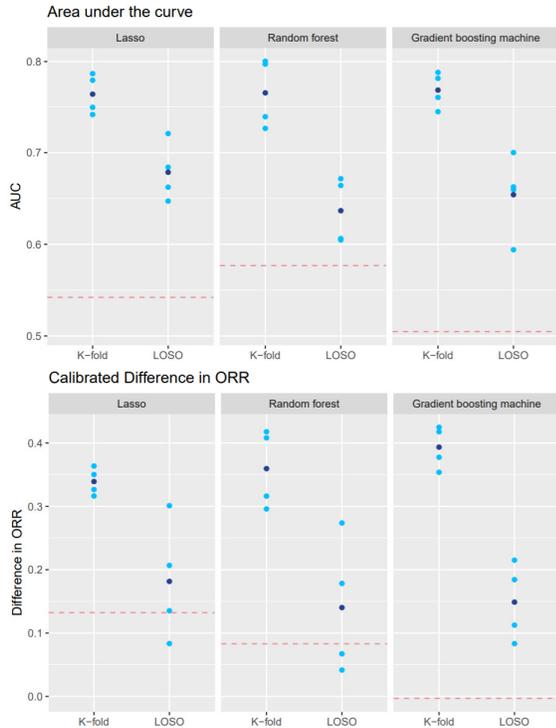

**Figure 1.** Dot plots of AUC (top) and ΔORR (bottom) estimates from K-fold vs LOSO cv using case study data, comparing estimates from each training fold (light blue dot) and the the mean (dark blue dot) to the true value from predictions about the "future" trial (red dashed line) with target prevalence of 50%

The conclusion held true when comparing the two cv methods with calibrated classification accuracy. The advantage of LOSO cv, as shown in all three cases, originates from its capability to address overfitting due to training scenarios and reduce overoptimism in assessing predictive power.

## 3 Simulation study

### 3.0.1 Design.
To further understand overfitting to training scenarios in different settings, we ran a series of simulation studies. We set up four legacy trials and one future trial each with a sample size of 500 patients. We simulated a causal feature $X \overset{i.i.d.}{\sim} N(0, 1)$, and defined a continuous outcome $Y = \beta X + \epsilon, \epsilon \overset{i.i.d.}{\sim} N(0, 1)$ where $\beta$ is a regression coefficient and $\epsilon$ is a randomly generated error term. Binary outcome was simulated through assigning each patient a value based on a probability $p = \frac{e^{\beta X}}{1+e^{\beta X}}$. It corresponds to the response status for each patient in the case study as well as in many real-world trials. For each trial, we further simulated 30 covariates following a multivariate normal distribution and randomly selected 15 of them to correlate with $X$ with correlation $\rho$. In this way, correlated features can vary across trials and the five trials have both shared and distinct features. 300 noise features were added to each trial.

The correlation ($\rho$) between covariates and the causal feature $X$ shifts across trials to account for heterogeneity across studies in the real world. $X$ represents the underlying biological mechanism that we don't get to measure directly and accurately, and $X$ remained unobserved when we built predictive models. The covariates represent the biomarkers collectable in each trial which could vary due to different assays used, different genes measured or other factors. Trials may have both shared and distinct covariates.

The performance of biomarker models were evaluated on simulated data with both binary and continuous outcomes. For the former, the same set of evaluation metrics was deployed as in the case study. For continuous outcomes, two cv methods was compared based on how close their estimation of the generalized $R^2$ was compared to the true value.

### 3.0.2 Results.
Figure 2 on the top shows the simulation results for binary outcomes when the model was evaluated with calibrated ΔORR. It confirmed the superiority of LOSO cv over K-fold cv in terms of managing the overoptimism when estimating effect size. We found the same trend when the model was evaluated using AUC and calibrated classification accuracy.

With continuous outcomes, the true generalized $R^2$ was much lower than what K-fold cv estimated. However, LOSO cv is able to yield an objective evaluation of the predictive power of the biomarkers. Such results stood for all three statistical models used: Lasso regression, random forest and gradient boosting machine (Figure 2). A series of sensitivity analysis was performed through tuning the parameters (sample size n, correlation $\rho$, effect $\beta$) and LOSO cv consistently outperformed K-fold cv and returned better estimation of generalization error of the biomarker. Additionally, we identified two simulations where we may expect more severe overestimation in generalized $R^2$ given by K-fold cv: (1) when we decreased the number of biomarkers correlated with the causal feature X (unobserved) in each trial to have fewer overlapping biomarkers and create larger heterogeneity across the simulated trials; (2) when we increased correlation $\rho$ to make biomarkers highly correlated with the unobserved causal factor. In both situations, the K-fold cv overestimates

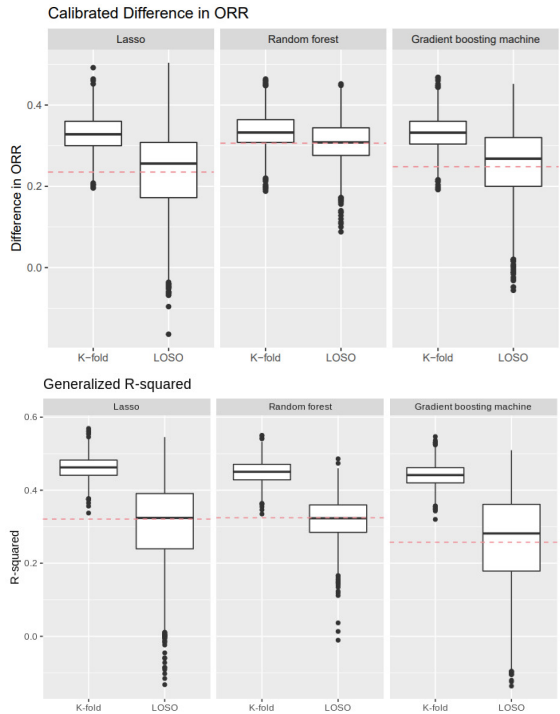

**Figure 2.** Box plots of ΔORR (top) and generalized $R^2$ (bottom) estimates from K-fold vs LOSO cv using simulation data, comparing ΔORR estimate (box) from each simulation to the true ΔORR (red dashed line) with target prevalence of 50%, and generalized $R^2$ estimate (box) from each simulation to the true generalized $R^2$ (red dashed line)

the true generalized $R^2$ to a larger extent, while LOSO cv corrects the overoptimism consistently.

## 4 Conclusion

Both the clinical trial case study and the simulation studies showed that LOSO cv compared to K-fold cv provides a more objective evaluation of biomarker's true effect size when predicted on future trials, particularly when studies are heterogeneous and when highly correlated candidate biomarkers are present. Due to the complexity of the case study data, LOSO cv did not approximate the performance of the biomarker exactly, yet it gave a much closer estimation than K-fold cv. When the data structure is simpler as demonstrated in the simulations, the estimation provided by LOSO cv closely aligned with the true effect size of the biomarker.

The superiority of LOSO holds across different statistical methods, changing association between biomarkers and outcomes, and varying correlation structure between studies. LOSO is able to outperform K-fold when the outcome is either continuous or binary outcomes, and such phenomenon is universal to a broad range of performance metrics.

## 5 Discussion

The principle behind LOSO cv, or any other "fit-for-purpose" cv is to perturb the randomness in traditional cv methods or bootstrapping. Instead of assessing the predictability of the biomarker on random splits of data which does not capture between-study heterogeneity, "fit-for-purpose" cv aims to set up training and testing folds in the same way as legacy and future trials in the real scenario. In our case study, the legacy trials differ in disease stage and the inclusion/exclusion criteria. In real biomarker development, we may know how future data will shift from legacy data and can take precautions to reduce overoptimism in biomarker's effect size by simulating pseudo data based on the structure shift and adding it in the cv process.

In this manuscript, we focused on examples in identifying patients with higher probability to respond to the treatment. This is a simple use case for biomarker development, but the idea of using "fit-for-purpose" cv to evaluate effect size can be applied when identifying a predictive biomarker or a prognostic biomarker using randomized data. Instead of using predictive accuracy as the summary statistics, one may consider the treatment difference between two arms in the selected population, the treatment-biomarker interaction p-value, or similar measures.

"Fit-for-purpose" cv provides us a way to objectively estimate the effect size of biomarkers. It's a great tool for evaluation, but does not build us a generalizable biomarker. Model generalizability remains a major challenge that still needs to be solved in biomarker development as well as in many other research fields (e.g. image and voice recognition). In some research fields, many novel machine learning or statistical methods have been studied to find generalizable signatures (e.g. adversarial training, domain adaptation, invariant risk minimization). Likewise, future efforts could also be directed towards advancing statistical models that can exploit the heterogeneity between trials and build a generalizable clinical biomarker.

# Appendix

## 1 ILLUSTRATION OF SAMPLE HETEROGENEITY ACROSS STUDIES

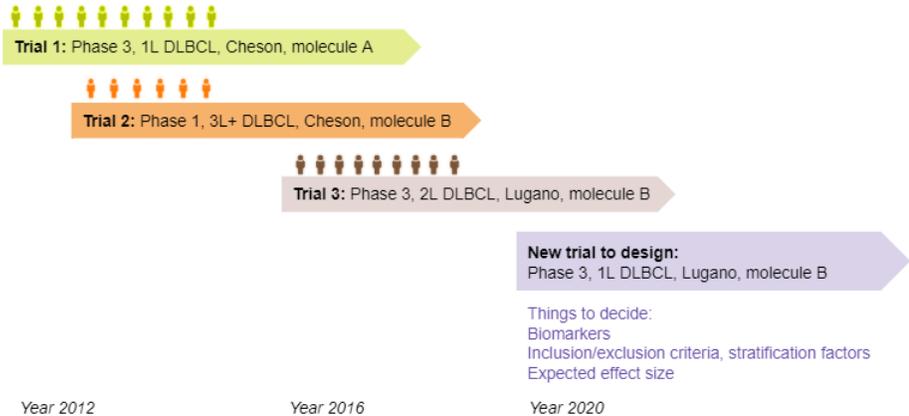

Fig. 1. Illustration of sample heterogeneity across studies, with three legacy trials and one future trial. This figure shows a simplified example about designing a first 1L DLBCL trial of molecule B. For the trial, we need to determine the enrollment criteria, choose the biomarker to collect, and predict treatment effects for sample size calculation purposes. The legacy data available may come from different indications (Trial 2 and 3), measure different molecules (Trial 1), and evaluate outcome using a different endpoint definition (Trial 1 and 2). With such heterogeneity across legacy trials and between legacy and future trials, the model is prone to overfit to the training "context" but yield suboptimal outcomes when applied to a new batch of data.

## 2 ILLUSTRATION OF TRAINING VS VALIDATION SETS SPLIT FOR K-FOLD CROSS-VALIDATION

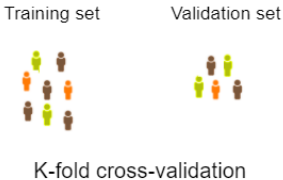

Fig. 2. Illustration of training vs validation sets split for K-fold cross-validation, with patients of the same color from the same trial. Conventional cross-validation methods such as K-fold randomly split data into separate training and validation sets so the biomarkers are tested on an "unseen" batch of data. When multiple legacy studies are available, researchers sometimes pool data from multiple studies. In this case, we can also apply K-fold cross-validation such that the training and validation parts have patients randomly chosen from all available studies.

## 3 ILLUSTRATION OF TRAINING VS VALIDATION SETS SPLIT FOR LOSO

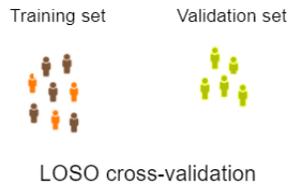

Fig. 3. Illustration of training vs validation sets split for LOSO, with patients of the same color from the same trial. To correct the over-optimism in our estimation, we recommend incorporating "fit-for-purpose" cross-validation, in which the training and validation folds are set up to mimic the difference between the learning data set and the future data set. For example, if we suspect that the future trial differs from legacy trials at study-level, we can use leave-one-study-out (LOSO) cross-validation which evaluates test error using unseen trials in the training dataset.

## 4 ILLUSTRATION OF K-FOLD CROSS-VALIDATION INCLUDING NESTED CROSS-VALIDATION

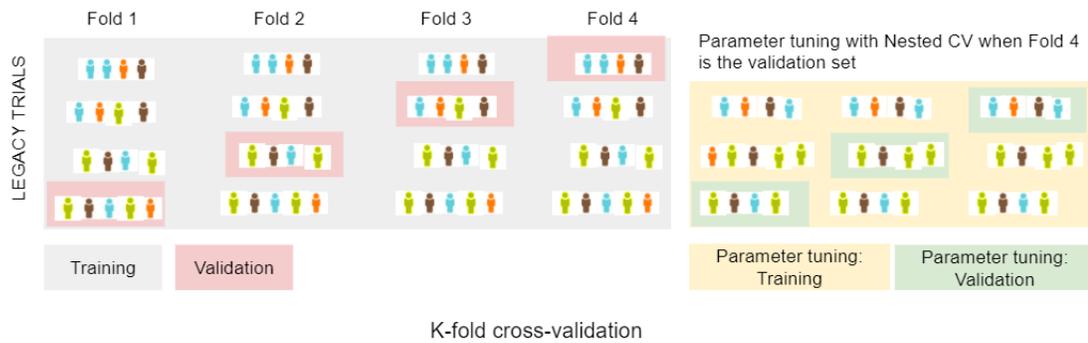

Fig. 4. Illustration of K-fold Cross-validation including nested cross-validation (left panel, main cross-validation process to estimate model performance; right panel: nested cross-validation to tune parameters in the model)

## 5 ILLUSTRATION OF LOSO CROSS-VALIDATION INCLUDING NESTED CROSS-VALIDATION

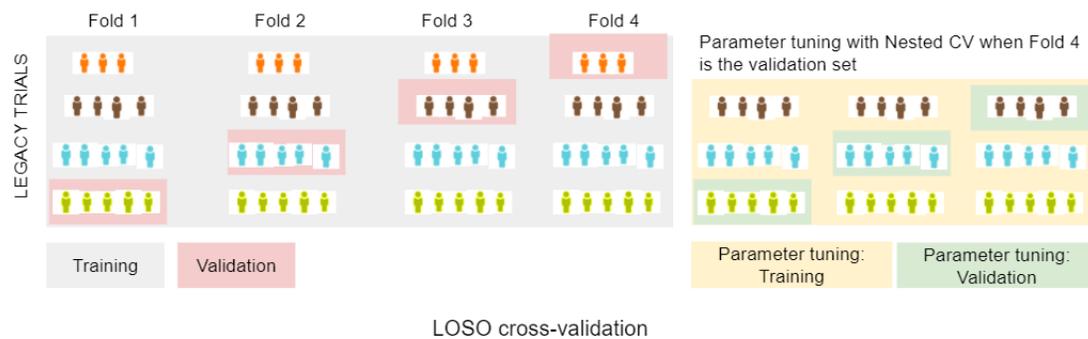

Fig. 5. Illustration of LOSO cross-validation including nested cross-validation (left panel, main cross-validation process to estimate model performance; right panel: nested cross-validation to tune parameters in the model)

## 6 STATISTICAL MODELS SET-UP, CASE STUDY

We fit Lasso regression to the case study data with penalty parameter $\rho$ chosen by nested cross-validation. During cross-validation, AUC was used to evaluate the performance of $\lambda$, and the best performer was used in the final model fitted on all legacy trials to predict future outcomes.

For random forests, we used a fixed number of 500 trees. Gini impurity was used to determine the best splits for each decision tree. Two hyperparameters were tuned using nested cross-validation: number of features to include at each node for splitting and the minimum internal node size to split on. AUC was also used to evaluate the performance of the two parameters.

In a gradient boosting machine, the maximum depth of each tree is fixed at 1 and the minimum number of observations in the terminal nodes of the trees is set at 10. Using nested cross-validation, we tuned two parameters: number of trees to fit and shrinkage (learning rate). Again, we calculated AUC to compare and select the best performing parameters.

## 7 ESTIMATED ΔORR BY K-FOLD VS LOSO CROSS-VALIDATION, CASE STUDY DATA

Table 1. ORR1 represents the ORR calculated among the biomarker positive group. ORR0 corresponds to the ORR calculated among the biomarker negative group. ΔORR = ORR1 - ORR0.

|        | Lasso |       |       | Random Forest |       |       | Gradient Boosting Machine |       |        |
|--------|-------|-------|-------|---------------|-------|-------|---------------------------|-------|--------|
|        | ORR1  | ORR0  | ΔORR  | ORR1          | ORR0  | ΔORR  | ORR1                      | ORR0  | ΔORR   |
| True   | 0.571 | 0.438 | 0.132 | 0.546         | 0.463 | 0.083 | 0.503                     | 0.506 | -0.003 |
| K-fold | 0.522 | 0.184 | 0.339 | 0.533         | 0.173 | 0.360 | 0.550                     | 0.156 | 0.393  |
| LOSO   | 0.405 | 0.223 | 0.182 | 0.384         | 0.244 | 0.140 | 0.389                     | 0.240 | 0.149  |

## 8 ADDITIONAL CALIBRATION ESTIMATES BY K-FOLD VS LOSO CROSS-VALIDATION, CASE STUDY DATA

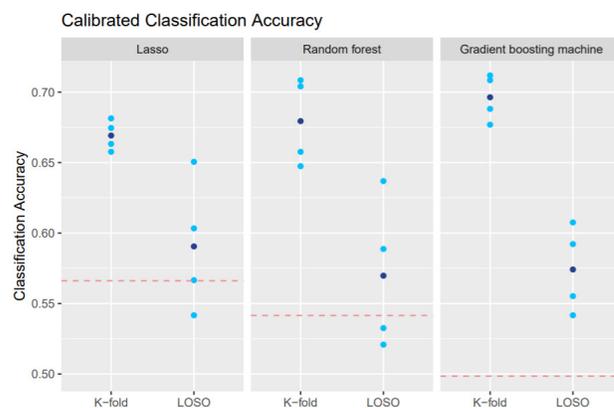

Fig. 6. Dot plot of calibrated classification accuracy estimates from K-fold vs LOSO cross-validation using case study data (left panel, Lasso regression; middle panel, random forest; right panel: gradient boosting machine), comparing calibrated classification accuracy from each training fold (light blue dot) with calibrated classification accuracy estimate summarized (dark blue dot) to the true AUC when the fitted model predicted the outcome of the "future" trial (red dashed line) with target prevalence of 50%.

## 9 UNCALIBRATED ESTIMATES BY K-FOLD VS LOSO CROSS-VALIDATION, CASE STUDY DATA

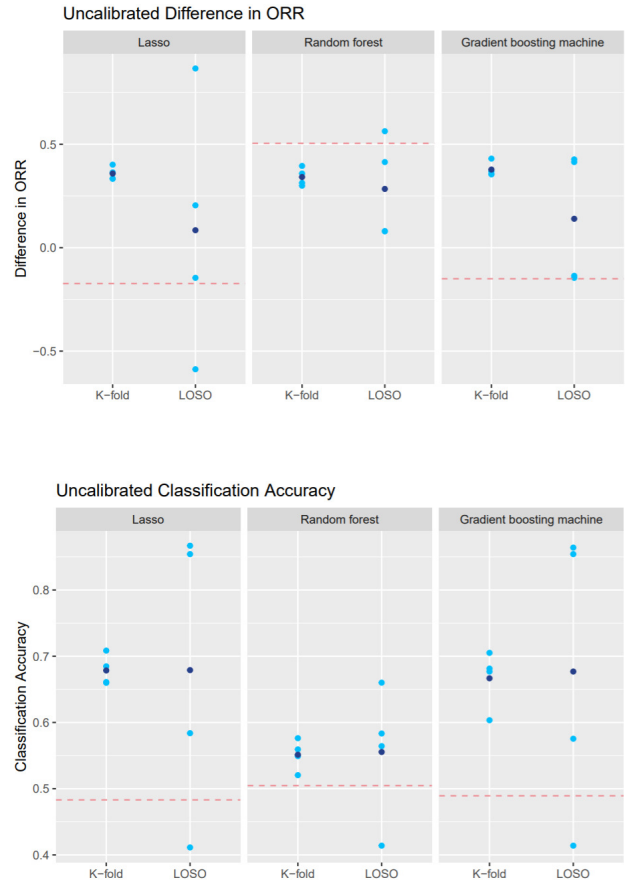

Fig. 7. Dot plot of uncalibrated estimates from K-fold vs LOSO cross-validation using case study data (top panel, ΔORR; bottom panel, classification accuracy), comparing uncalibrated estimates from each training fold (light blue dot) with uncalibrated estimates summarized (dark blue dot) to the true uncalibrated estimates when the fitted model predicted the outcome of the "future" trial (red dashed line) with target prevalence of 50%.

## 10 STATISTICAL MODEL SET-UP, SIMULATION STUDY

We simulated the trial data using the following steps.

(1) Simulated 4 legacy trials and 1 future trial with sample size of 500 each. For each trial, we had one causal feature X following a standard normal distribution.
(2) Simulated the continuous outcome $Y = X + \epsilon, \epsilon \overset{i.i.d.}{\sim} N(0, 1)$. For binary outcomes, we simulated from a Bernoulli outcome with probability
(3) Additionally, simulated 30 candidate biomarkers $Z_1..Z_{30}$ following N(0, 1). We randomly selected 15 of them to correlate with X with correlation $\rho$. The other 15 covariates were independent from X. For example, in one set of simulation with 4 legacy trials and 1 future trial, there was 1 correlated covariate overlapping in all 5 trials, 3 correlated covariates overlapping in 4 trials, 13 correlated covariates overlapping in 3 trials, 7 correlated covariates overlapping in 2 trials, and 5 correlated covariates that only exist in one trial.
(4) Lastly, simulated 300 noise features with a standard normal distribution.

When training the biomarkers using simulated legacy trials, we performed nested cross-validation to select parameters for the three statistical models: Lasso regression, random forest, and gradient boosting machine. For continuous outcomes, we tuned the penalty parameter $\lambda$ in Lasso regression and used MSE to evaluate the performance of $\lambda$ in nested cross-validation. For random forest, we also used a fixed number of 500 trees and fixed the minimum internal node size to split on to be 5. Variance was used to determine the best splits for each decision tree. We used nested cross-validation to tune the number of covariates to include in the tree at each node and this parameter was evaluated by $R^2$. For the gradient boosting machine model, the maximum depth of each tree is fixed at 1 and the minimum number of observations in the terminal nodes of the trees is set at 10. Using nested cross-validation, we tuned two parameters: number of trees to fit and shrinkage. Again, we calculated $R^2$ to compare and select the best performing parameters.

The model set-up for binary outcome remained the same as in the case study.

## 11 AUC ESTIMATES FROM K-FOLD VS LOSO CROSS-VALIDATION, SIMULATION STUDY

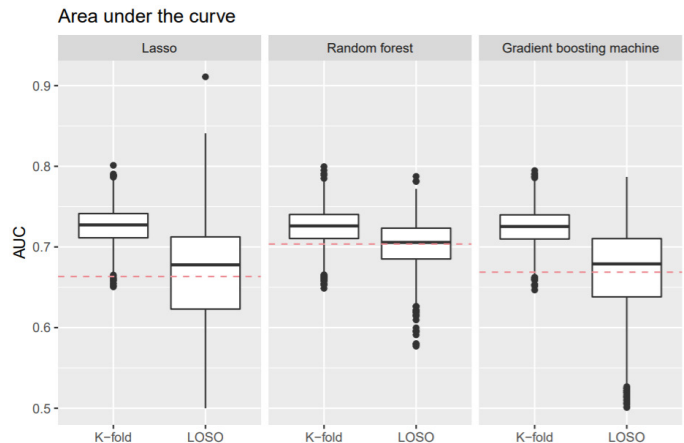

Fig. 8. Box plot of AUC estimates from K-fold vs LOSO cross-validation using simulation study data (left panel, Lasso regression; middle panel, random forest; right panel: gradient boosting machine), comparing AUC estimate from each simulation to the true AUC when the fitted model predicted the outcome of the future trial (red dashed line). While the LOSO closely approximates the true performance of the biomarker in the future trial, K-fold cross-validation overestimates the effect size.

## 12 ADDITIONAL CALIBRATION ESTIMATES BY K-FOLD VS LOSO CROSS-VALIDATION, SIMULATION STUDY

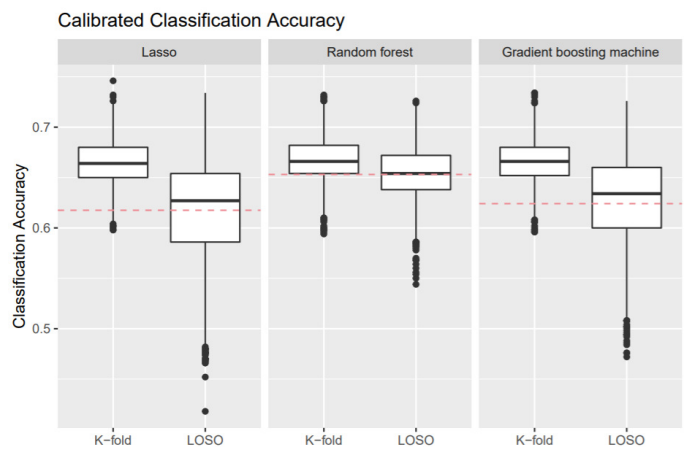

Fig. 9. Box plot of calibrated classification accuracy estimates from K-fold vs LOSO cross-validation using simulation study data (left panel, Lasso regression; middle panel, random forest; right panel: gradient boosting machine), comparing classification accuracy estimate from each simulation to the true classification accuracy when the fitted model predicted the outcome of the future trial (red dashed line) with target prevalence of 50%.

## 13 UNCALIBRATED ESTIMATES BY K-FOLD VS LOSO CROSS-VALIDATION, SIMULATION STUDY

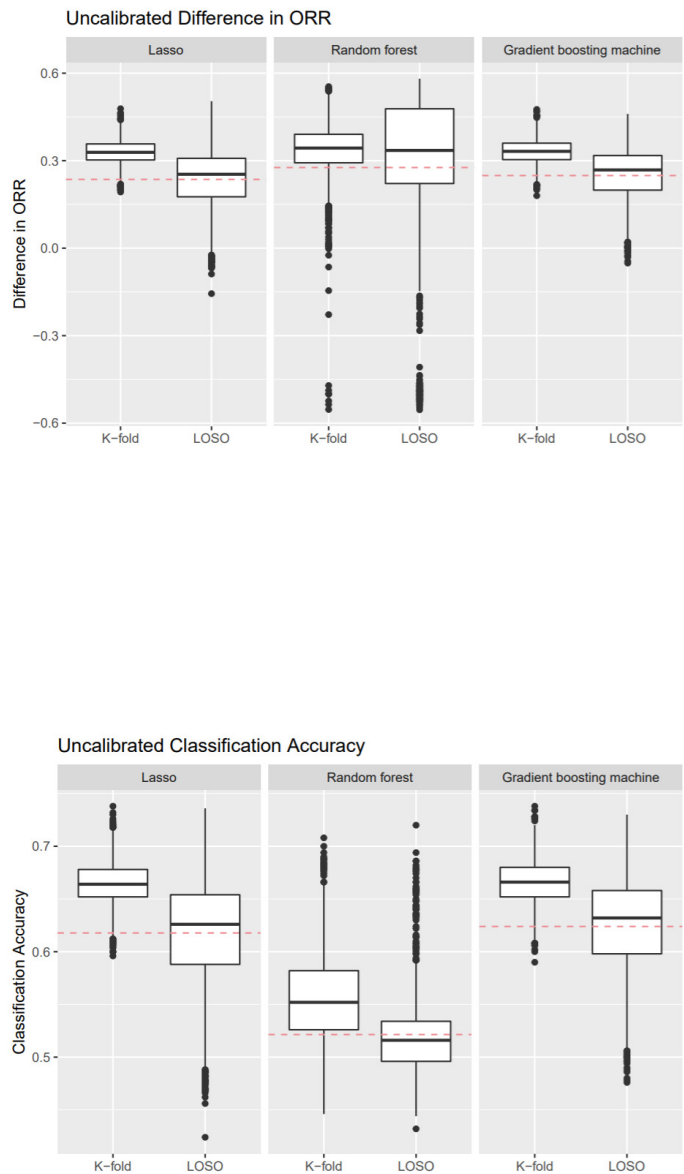

Fig. 10. Box plot of uncalibrated estimates from K-fold vs LOSO cross-validation using simulation study data (top panel, ΔORR; bottom panel, classification accuracy), comparing uncalibrated estimates estimate from each simulation to the true uncalibrated estimates when the fitted model predicted the outcome of the future trial (red dashed line) with target prevalence of 50%.

## 14 GENERALIZED $R^2$ ESTIMATES FROM K-FOLD VS LOSO CROSS-VALIDATION, SIMULATION STUDY

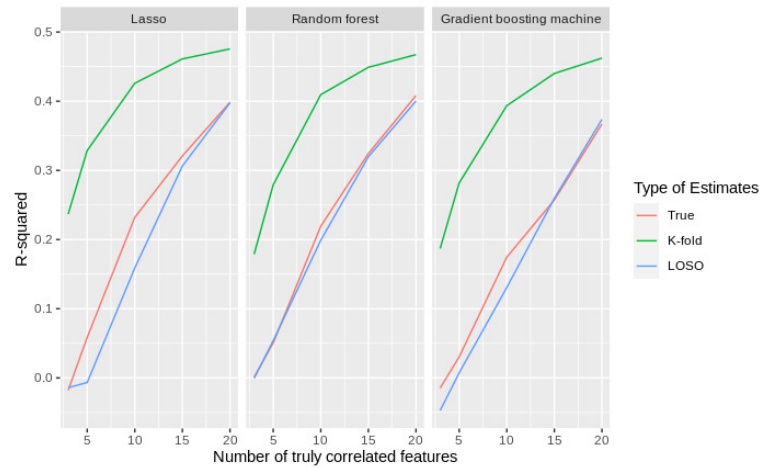

Fig. 11. Line chart of generalized $R^2$ estimates from K-fold vs LOSO cross-validation using simulation study data (left panel, Lasso regression; middle panel, random forest; right panel: gradient boosting machine). The figure shows the trend of generalized $R^2$ by number of correlates ($\beta$ = 1, $\rho$ = 0.9, 300 noise features), comparing mean generalized $R^2$ estimate from K-fold cross-validation (green line) and generalized $R^2$ estimate from LOSO cross-validation (blue line) to the true generalized $R^2$ when the fitted model predicted the outcome of the future trial (red line).

## 15 GENERALIZED $R^2$ ESTIMATES FROM K-FOLD VS LOSO CROSS-VALIDATION, SIMULATION STUDY WITH HIGHLY CORRELATED BIOMARKERS

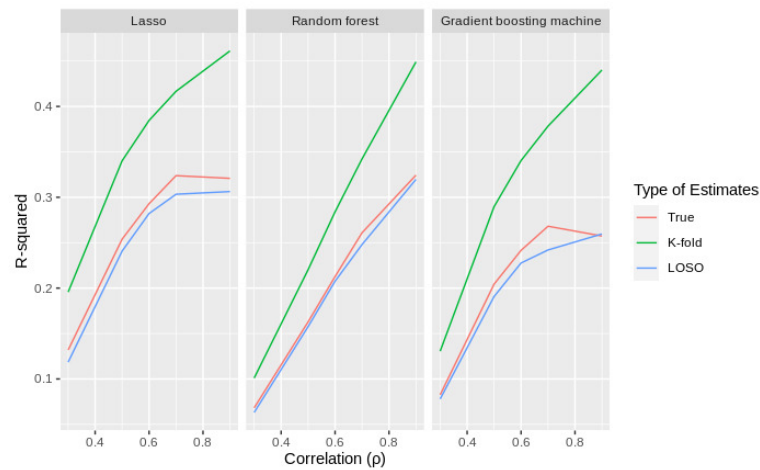

Fig. 12. Line chart of generalized $R^2$ estimates from K-fold vs LOSO cross-validation using simulation study data (left panel, Lasso regression; middle panel, random forest; right panel: gradient boosting machine). The figure shows the trend of generalized $R^2$ by correlation ($\beta$ = 1, 15 correlates, 300 noise features), comparing mean generalized $R^2$ estimate from K-fold cross-validation (green line) and generalized $R^2$ estimate from LOSO cross-validation (blue line) to the true generalized $R^2$ when the fitted model predicted the outcome of the future trial (red line). In the situations where highly correlated features exist in the data, the K-fold cross-validation overestimates the predictive performance of biomarkers more severely, and the LOSO cross-validation corrects the overoptimism.

## 16 OVERESTIMATION OF GENERALIZED $R^2$ WITH INCREASING CAUSAL FEATURE EFFECT

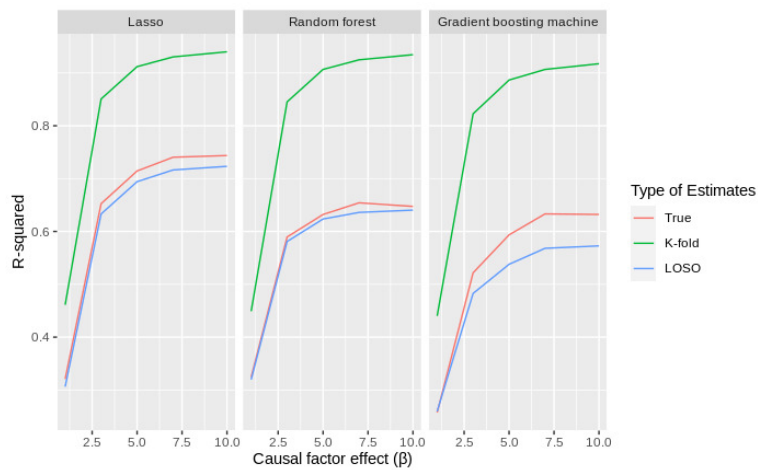

Fig. 13. Line chart of generalized $R^2$ estimates from K-fold vs LOSO cross-validation using simulation study data (left panel, Lasso regression; middle panel, random forest; right panel: gradient boosting machine). The figure shows the trend of generalized $R^2$ by causal effect $\beta$ ($\rho$ = 0.9, 15 correlates, 300 noise features), comparing mean generalized $R^2$ estimate from K-fold cross-validation (green line) and generalized $R^2$ estimate from LOSO cross-validation (blue line) to the true generalized $R^2$ when the fitted model predicted the outcome of the future trial (red line).

## 17 DESCRIPTION OF SOURCE CLINICAL DATA

Clinical data from 4 different Roche clinical trials (POPLAR, OAK, IMpower150, and IMpower131) were harmonized, and only cohorts with atezolizumab treatment were used in our analyses. The harmonisation was undertaken to the latest Study Data Tabulation Model (SDTM) version and standard analysis datasets were generated with re-derivation of endpoints, which may not necessarily be the same as in the Clinical Study Report (CSR). Legacy studies POPLAR (144 patients in atezolizumab cohort with clinical cutoff of Apr 07, 2017), OAK (613 patients in atezolizumab cohort with clinical cutoff of Jan 23, 2017), and IMpower150 (402 patients in atezolizumab plus carboplatin plus paclitaxel cohort and 400 patients in atezolizumab plus carboplatin plus paclitaxel plus bevacizumab cohort with clinical extract of Nov 18, 2019) were used as our training data, and IMpower131 atezolizumab cohorts (338 patients in atezolizumab plus carboplatin plus paclitaxel cohort) with a clinical data extract of Nov 18, 2019 were used as our testing data to predict responses separately. For each study, Illumina RNAseq assay data for measuring gene expression of up to 56,494 genes has been integrated with the clinical data. After data processing of removing patients with missing baseline data and/or RNA-seq data, our sample size of the training cohorts is 1,179, and the sample sizes are 320 for atezolizumab plus carboplatin plus paclitaxel cohort. Responder is defined as patients with their best confirmed overall response by investigator of Complete Response (CR) or Partial Response (PR), and non-responder otherwise.

### ACKNOWLEDGEMENT

This research was supported by Genentech Inc.. We would like to acknowledge Rob Tibshirani, Svetlana Lyalina, Ariel Lopez Chavez, Nick Bulley, and Jerry Hsu for their inputs during Roche Advanced Analytics Data Challenge 2 which motivated the development of this manuscript.

## DATA AVAILABILITY STATEMENT

For the use case, the clinical data are available at:

OAK: https://search.vivli.org/studyDetails/b455270d-39cd-4a6f-95c9-a2d4187f73bd

POPLAR: https://search.vivli.org/studyDetails/2169c08e-5e6f-46c0-9c33-d620867edc35

Impower150: https://search.vivli.org/studyDetails/6e7c8826-a4e3-42e4-b2c1-37eaecc94122

Impower131: https://search.vivli.org/studyDetails/d71bf0ce-bef3-4133-9532-add6df1828ae

POPLAR and OAK RNAseq data was submitted to EGA: EGAS00001005013